\documentclass[onecolumn,amsmath,amssymb]{qm2008_abs}
\usepackage{graphicx}
\usepackage{dcolumn}
\usepackage{bm}
\def\piz{$\pi^{0}$ }
\def\et{$\eta$ }
\def\etapi{$\eta/\pi^{0}$ }
\def\raa{$R_{\mathrm{AA}}$ }
\def\pt{$p_{\mathrm{T}}$ }
\def\roots{$\sqrt{s_{\mathrm{NN}}}$ }

\begin{document}

\title{{\Large Measurement of \piz and \et Mesons with PHENIX in
    \roots = 200~GeV Au+Au Collisions at RHIC}} 

\bigskip
\bigskip
\author{\large Baldo Sahlm\"uller for the PHENIX collaboration}
\email{sahlmul@uni-muenster.de}
\affiliation{Institut f\"ur Kernphysik, University of M\"unster, M\"unster, Germany}
\bigskip
\bigskip

\begin{abstract}
\leftskip1.0cm
\rightskip1.0cm
The \piz meson has been a crucial proble for observing jet quenching
in ultrarelativistic heavy-ion collisions at RHIC. Measurements of the
\et meson in the same collisions have also shed light on a possible
dependence of the observed suppression on the particle species. The
preliminary \piz nuclear modification factor \raa from the 2004 RHIC
run allowed a first systematic comparison between a precise
measurement with high statistics and theoretical calculations,
constraining model parameters such as the initial gluon density
d$N^{g}/$d$y$, and the transport coefficient $\hat{q}$. The final \piz
spectra and \raa are shown as well as the first \et results obtained
with both PHENIX electromagnetic calorimeters.
\end{abstract}

\maketitle

\section{Introduction}
Previous measurements at RHIC have shown a significant suppression of
\piz, \et, and charged hadrons in central Au+Au collisions at \roots 
= 200 GeV compared to binary scaled p+p collisions
\cite{pi0suppr,suppr_star}. The $\sim$ 30 times larger dataset from
the 2004 RHIC run allows a more precise measurement of the observed
suppression, reaching higher transverse momenta as well. Both the
higher precision and the higher \pt reach are crucial for techniques
used to constrain parameters of theoretical models such as the initial
gluon density d$N^{g}/$d$y$, and the transport coefficient $\hat{q}$
with the measured data \cite{ppg079}. The data can also be used to
calculate the dependence of the suppression on the number of participant
nucleons $N_{\mathrm{part}}$ \cite{ppg080}. The measurement of the \et meson up
to higher transverse momenta than in previous measurements
sheds further light on the understanding of energy loss. Together with
the measurement of direct photons and other hadrons, the PHENIX data
allow more detailed studies of predictions and assumtions of parton
energy loss models.

\section{Measurement of \piz and \et Spectra}

\begin{figure}[t]
\begin{minipage}[t]{60mm}
\includegraphics[height=3in]{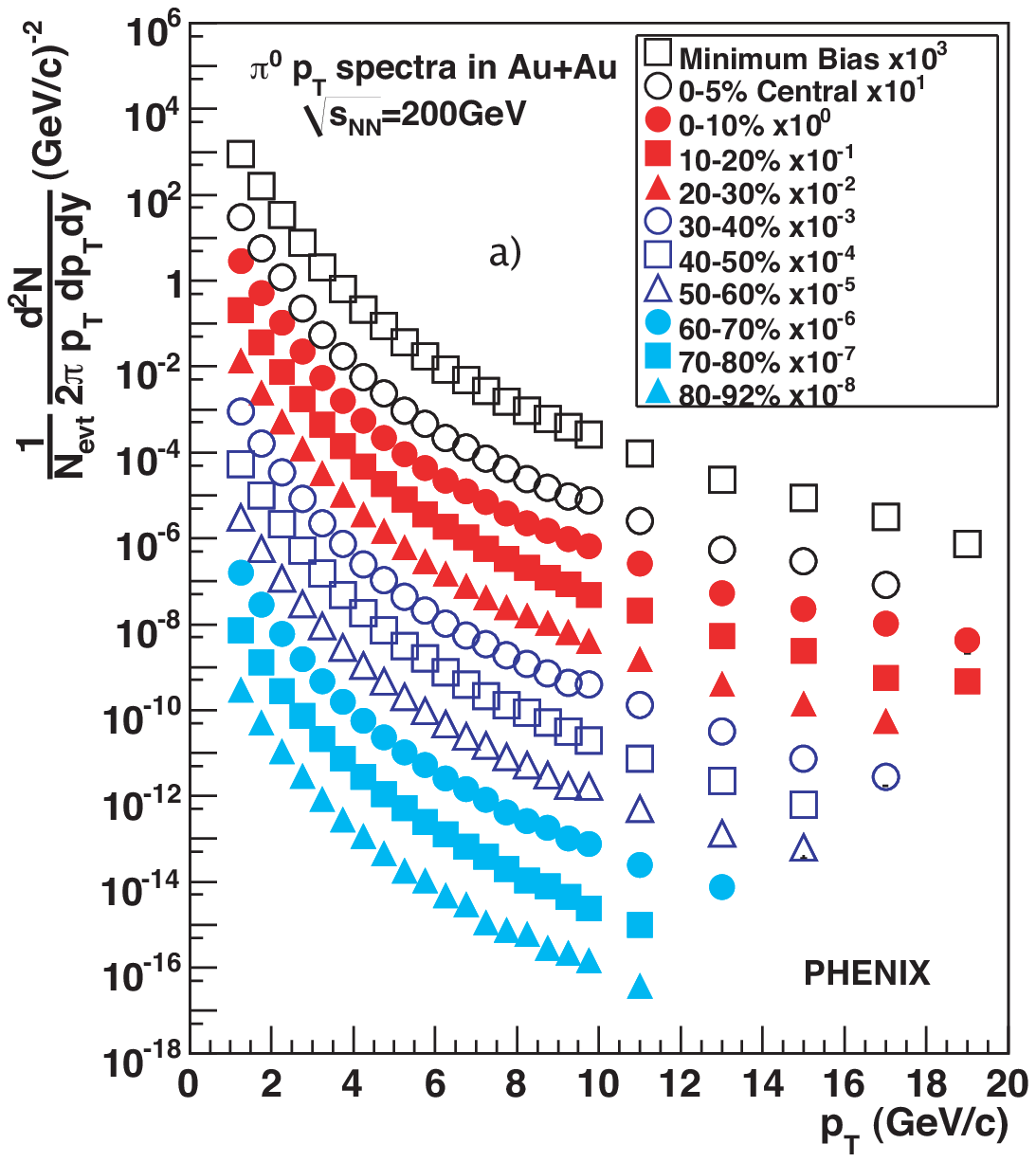}
\end{minipage}
\hspace{1cm}
\begin{minipage}[b]{60mm}
\includegraphics[height=3in]{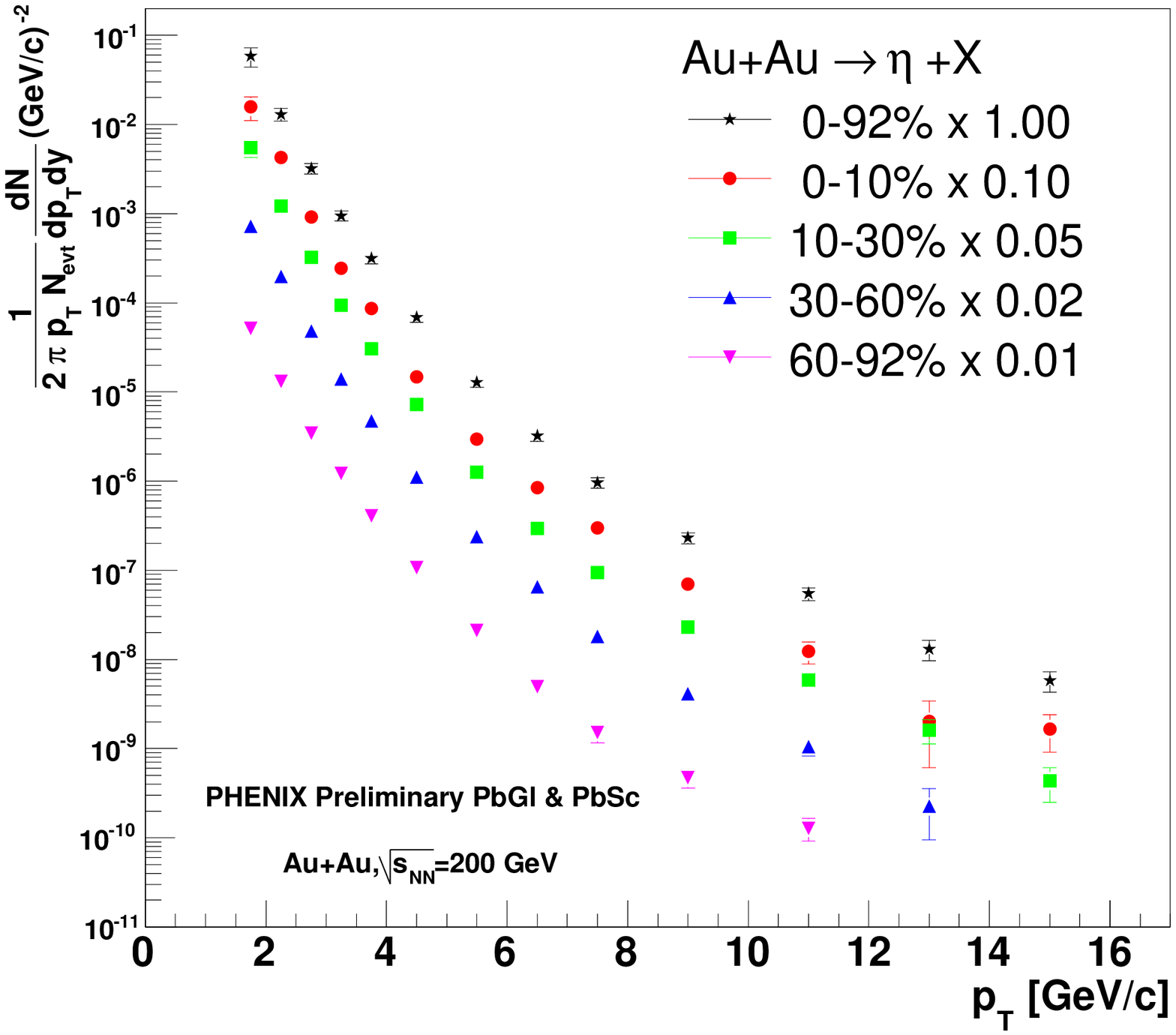}
\end{minipage}
\caption{\label{fig:yields}Fully corrected invariant yields for a)
  \piz's \cite{ppg080} and b) \et's
  for different centrality selections in Au+Au collisions at \roots =
  200 GeV. The error bars show the \pt uncorrelated errors.}
\end{figure}

Both neutral mesons \piz and \et are measured in the PHENIX
experiment via their two photon decay \cite{ppg054, ppg055}. The decay
photons are measured
with the Electromagnetic Calorimeter, consisting of six sectors of
lead scintillator sandwich calorimeters and two sectors of lead glass
calorimeters, at midrapidity. Each sector covers 22.5 degrees in
azimuth, leading to a total coverage of 180 degrees in azimuth and
$(|\eta| < 0.35)$ in pseudorapidity. The centrality of a collision is
determined with the correlation of the signals in the Beam-Beam
Counters (BBC) and the Zero Degree Calorimeters (ZDC).\\
Uncorrected particle yields are
extracted with an invariant mass analysis using event mixing for the
substraction of background. Therefore in a first step
the invariant mass of all photon candidates in one event is
calculated. This method leads to a large combinatorial background.
To estimate this background, the invariant mass of all photon
candidates in the current event with all photon candidates in one or
more other events is calculated. This mixed events invariant mass
distribution is finally scaled to the background in the real events
distribution outside the \piz or \et peak, subtracted from this real
event invariant mass distribution, and finally the peak is
integrated. These raw yields are then
corrected for different effects to obtain the physical particle
spectra. Corrections include the geometric acceptance, the
reconstruction efficiency, the conversion of decay photons, the
merging of \piz decay photon clusters at high transverse momenta, and
a correction for the center of the \pt bins taking into account the
local slope.\\
Results are obtained for both detectors indepentently and finally
combined, using the statistical and systematic uncertainties as a
weight. For the \et, this measurement is the first one using 
the PbGl detector in Au+Au collisions. The fully corrected invariant
yields for both particles are shown in Fig. \ref{fig:yields}.\\

\section{$R_{\mathrm{AA}}$}

\begin{figure}[t]
\begin{minipage}[t]{70mm}
\includegraphics[height=3in]{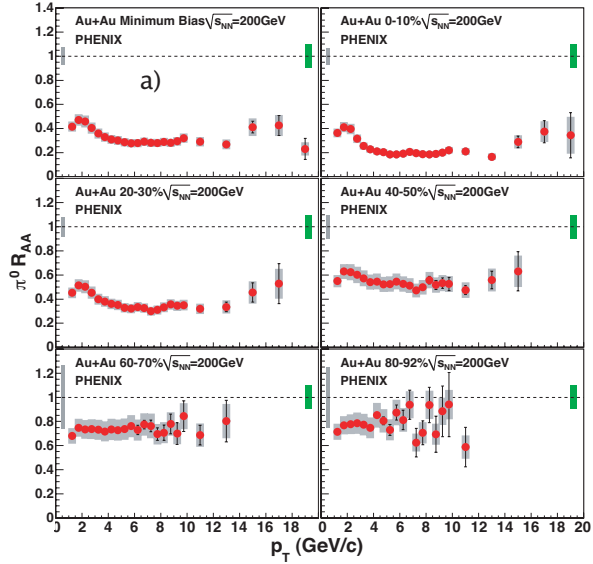}
\end{minipage}
\hspace{\fill}
\begin{minipage}[b]{65mm}
\includegraphics[height=2in]{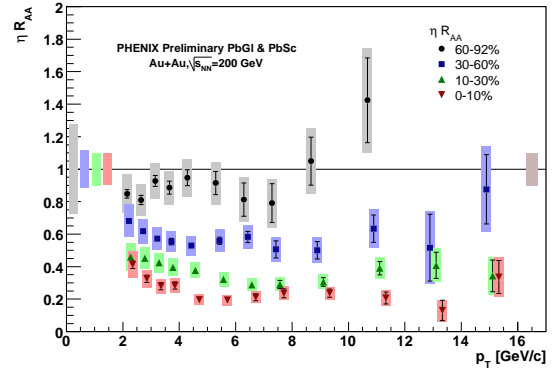}
\caption{\label{fig:raa}Nuclear modification factor \raa for a) \piz
  \cite{ppg080} and b) \et for different centrality selections in
  Au+Au collisions at \roots = 
  200 GeV. The error bars show the \pt uncorrelated errors, the boxes
  around the points show the \pt correlated errors, the box at the
  left shows the normalization uncertainty.}
\end{minipage}
\end{figure}

The nuclear modification factor \raa is a measure to compare the
particle spectra in nucleus nucleus collisions with the ones in p+p
collisions. It is defined as $R_{\mathrm{AA}} =
\frac{1/N_{\mathrm{evt}}d^2N/dp_{\mathrm{T}}dy|_{\mathrm{A+A}}}
{T_{\mathrm{AA}}d^2\sigma/dp_{\mathrm{T}}dy|_{\mathrm{p+p}}}$,
with $T_{AA}$ denoting the nuclear overlap function. It is shown in
Fig. \ref{fig:raa}a for neutral pions in 
Au+Au collisions at \roots = 200~GeV for five different centrality
selections and 0-92\%. A clear centrality dependence is
observed, \raa decreases towards central events, reaching
$\sim$~0.2 in central events. \raa is almost constant up to
\pt~$\approx$~20~GeV/$c$ in all centralities. For the \et meson, \raa
is shown in Fig. \ref{fig:raa}b for four different centrality
selections. The same centrality dependence is seen here, and the \et
is suppressed by a factor of $\sim$~5 in central events similar to the \piz.

\section{\etapi Ratio}

\begin{figure}[b]
\begin{minipage}[b]{90mm}
\includegraphics[height=2.5in]{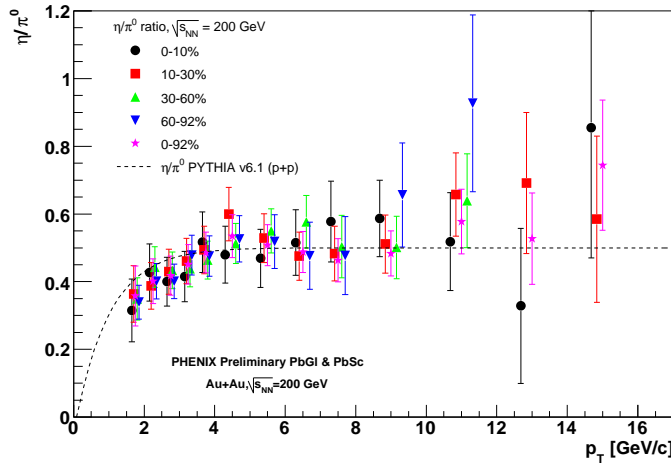}
\end{minipage}
\begin{minipage}[b]{45mm}
\caption{\label{fig:etapi}Ratio of \et and \piz in Au+Au at
  \roots = 200 GeV for different centrality selections and minimum
  bias in comparison with a PYTHIA \cite{pythia} calculation. The
  error bars show the \pt uncorrelated errors.}
\end{minipage}
\end{figure}

The \etapi ratio for
Au+Au collisions at \roots = 200~GeV for different centrality
selections is shown in Fig. \ref{fig:etapi} in comparison with a
PYTHIA \cite{pythia} calculation for p+p collisions. The ratio is found to
be independent of centrality over the whole \pt range and the PYTHIA
curve is in good agreement with the measured data. When fit with a
constant for \pt $>$ 2~GeV/$c$, the parameter varies from $c=0.471 \pm
0.028$ (stat) for 0-10\% to $c=0.462 \pm 0.023$ (stat) for 60-92\%. The
ratio is consistent with
data from earlier measurements at different energies and collision
systems \cite{ppg055}. A possible explanation is that
the suppression of high-\pt hadrons occurs at the partonic level and 
that the fragmentation is not affected by the medium.

\section{Comparison with other Particles}

\begin{figure}[t]
\begin{minipage}[t]{90mm}
\includegraphics[height=2.5in]{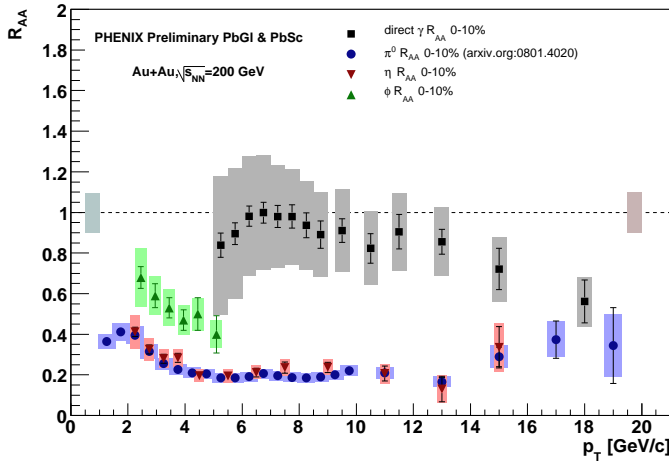}
\end{minipage}
\begin{minipage}[b]{50mm}
\caption{\label{fig:raa_all}Nuclear modification factor \raa for \piz
  and \et compared with \raa for direct photons and the $\phi$
  meson in 0-10\% most central Au+Au collisions at \roots=200~GeV. The
  error bars show the \pt uncorrelated errors, the boxes
  around the points show the \pt correlated errors, the boxes at the
  left show the normalization uncertainties (left: from the
  $T_{\mathrm{AA}}$ calculation, right: the p+p normalization
  uncertainty).}
\end{minipage}
\end{figure}

Fig. \ref{fig:raa_all} shows the nuclear modification factor for \piz
and \et in 0-10\% most central Au+Au collisions at \roots=200~GeV
together with \raa for direct photons and $\phi$ mesons. The same
suppression pattern of \piz and \et can clearly be seen, both mesons
are similarly suppressed up to \pt = 15~GeV/$c$. In contrast, direct
photons are not significantly suppressed up to the same
$p_{\mathrm{T}}$, but show an indication for suppression at the
highest \pt which is consistent with initial state effects
\cite{tadaaki}. The $\phi$, along with other mesons \cite{klaus}, is
less suppressed than the \piz and  the et. This interesting
observation might provide an important test for jet quenching models.

\section{Summary}

The PHENIX experiment has measured \piz and \et mesons in Au+Au
collisions at \roots = 200~GeV. The nuclear modification factor \raa
has been calculated, showing a similar suppression of both particles
up to high transverse momenta. Both mesons are suppressed by a factor
of $\sim$~5 in 0-10\% central Au+Au collisions. The production ratio
\etapi has also been calculated. It is centrality independent and
also independent of the collision system and energy which supports the
assumption of partonic energy loss. A comparison of \raa with direct
photons and $\phi$ mesons shows a different behaviour for both 
particles. While the direct-photon data are consistent with initial
state effects, the behaviour of the $\phi$ meson remains an
interesting open question.\\

\end{document}